\documentstyle[prl,aps,epsf,twocolumn]{revtex}

\begin{document}
\draft

\title{Magnetic Field Dependence of Magic Numbers in Small Quantum Dots}
\author{W.D. Heiss$^{*\dagger }$ and R.G. Nazmitdinov
$ ^{*\ddagger}$} \address{
$^*$ Institute for Nuclear Theory, University of Washington, Box 351550,
 Seattle, WA 98195, USA\\
$^{\dagger }$Department of
Physics, University of the Witwatersrand, Johannesburg, South Africa\\
$^{\ddagger}$ Joint Institute for Nuclear Research, Dubna, Russia} 
 \maketitle

\begin{abstract}
It is argued that various kind of shell structure which occurs at 
specific values of the magnetic field should be observable in small quantum dots
in the addition energies and the magnetic susceptibility.
\end{abstract}


Quantum dots are ideal mesoscopic objects for the
study of quantum mechanical properties and comparison with classical
behavior \cite{Kas}. The smaller the size of quantum dots, the larger
the prevalence of quantum effects upon the static and dynamic properties
of these systems. Quantum dots may be considered as true artificial atoms the
properties of which can be controlled by men. 

An important feature of finite Fermi systems is the existence of shells which
give rise to magic numbers. The periodic table of chemical elements is a
striking example for the dramatic difference between closed shell 
and partially filled shell systems; for nuclei and metallic clusters see
for example \cite {BM,NR,HB}. In recent experiments \cite{e1,e2,e3}
shell structure phenomena have been observed clearly for quantum dots.
In particular, the energy needed to place the extra electron (addition energy)
into a vertical quantum dot has characteristic maxima which correspond to 
the sequence of magic numbers of a two-dimensional harmonic
oscillator \cite{e2}. In this Letter we propose manifestations of shell 
effects that should be observable in the addition energy and the magnetic 
susceptibility for small quantum dots under the influence of a magnetic field.

We choose as the confining mean field for the electrons in quantum dots
the harmonic oscillator potential.
Any smooth and finite potential admitting bound states can, for the 
lowest few levels, generically be approximated by
the harmonic oscillator potential \cite{DE}. 
For small electron numbers the harmonic oscillator has been used successfully 
as a phenomenological effective confining potential in quantum dots
\cite{HK}. As was shown in the simple case of 
a two-electron system, the external
parabolic potential and the Coulomb interaction leads to a new
{\it oscillator} frequency of the effective mean field \cite{DN}. The
effect of an external homogeneous magnetic field can be calculated exactly 
for a three dimensional (3D) harmonic oscillator potential 
irrespective of the direction of the field \cite{HeNa97}. 
Our discussion here is based upon the 2D version of the Hamiltonian
\cite{HeNa97}. The magnetic field acts perpendicular to the plane of motion 
of the electrons and the spin degree of freedom is incorporated, i.e. 
 $H=\sum_{j=1}^N h_j$ 
where
\begin{equation}
h={1\over 2m^*}(\vec p-{e\over c}\vec A)^2+{m^*\over 2}(\omega _x^2 x^2
+\omega _y^2 y^2) + \mu^*\sigma _zB_z.           \label{ham}
\end{equation}
Here $\vec A=[\vec r \times \vec B]/2$ and $\sigma _z$ is the Pauli matrix.

\vspace*{-0.1cm}
\begin{figure}
\epsfxsize=2.5in
\centerline{ \epsffile{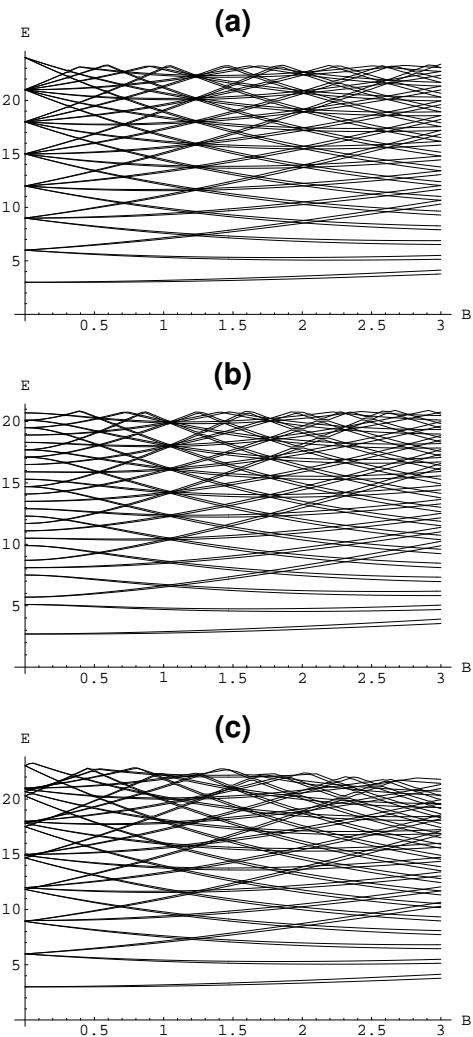}  }
\vglue 0.2cm
\caption{Single-particle spectra as a function of the magnetic field strength.
Spectra are displayed for (a) a plain isotropic two dimensional oscillator, 
(b) a deformed and (c) an isotropic oscillator including an $L^2$-term. For
better illustration the value $2\mu ^*$ is used for the spin magnetic
splitting in all Figures.}
\label{fig1}
\end{figure}

We do not take into account the effect of finite temperature; this is 
appropriate for experiments which are performed at temperatures  
$kT\ll \Delta$ with $\Delta $ being the mean level spacing. In the following
we use meV for the energy and Tesla for the magnetic field strength.

In Fig.1a we display the single-particle spectrum for the isotropic case
($\omega _x=\omega _y$) 
as a function of the magnetic field strength $B$ . The effective mass 
which determines the orbital magnetic moment for the electrons is chosen as 
$m^*=0.067m_e$. It leads to $\mu _B^{{\rm eff}}=15\mu _B$ while the effective 
spin magnetic moment is $\mu ^*=0.5\mu _B$. The 
magnetic orbital effect is much enhanced in comparison with the magnetic spin 
effect, yet the tiny spin splitting does produce signatures as we see below.

Shell structure occurs whenever the ratio of the two eigenmodes
$\Omega _{\pm }$ of the Hamiltonian (\ref{ham}) (see Ref.\cite{HeNa97})
is a rational number with a small numerator and denominator.
 The shell structure is particularly pronounced if 
the ratio is equal to one (for $B=0$ and $\omega _x=\omega _y$) or two (for 
$B\approx 1.23$ ) or three (for $B\approx 2.01$ ) and lesser
pronounced if the ratio is 3/2 (for $B=0.72$) or 5/2 (for $B=1.65$).
The values given here for $B$ depend on $m^*$ and $\omega _{x,y}$. 
As a consequence, a material with an even smaller effective mass $m^*$ would
show these effects for a correspondingly smaller magnetic field.
The magic numbers (including spin) turn out, for $B=0$, to be the 
usual sequence of the two dimensional isotropic oscillator, that is 
$2,6,12,20,\ldots $ \cite{e2}. For $B\approx 1.23$
we find a new shell structure {\em as if} the confining potential would be a 
deformed harmonic oscillator without magnetic field. The magic numbers are 
$2,4,8,12,18,24,\ldots $ which are just the numbers obtained from the two
dimensional oscillator with $\omega _>=2\omega _<$ ($\omega _>$ and 
$\omega _<$ denote the larger and smaller value of the two frequencies). 
Similarly, we get for $B\approx 2.01$ the magic numbers 
$2,4,6,10,14,18,24,\ldots $ which corresponds to $\omega _>=3\omega _<$.

If we start from the outset with a deformed mean field, i.e.~if we choose, 
say, $\omega _x=(1-\beta )\omega _y$ with $\beta >0$
\cite{HHW}, two major effects are found:
(i) the degeneracies (shell structure) are lifted at $B=0$ depending on the 
actual value of $\beta $, and (ii) the values for $B$ at which the new shell
structures occur are shifted to lower values. In Fig.1b we display an example
referring to $\beta =0.2$. The significance of this finding lies in the
restoration of shell structures by the magnetic field in an isolated quantum dot that
does not give rise to magic numbers at zero field strength due to deformation.
We mention that the choice $\beta =0.5$ would shift the pattern found at 
$B\approx 1.23$ in Fig.1a to the value $B=0$.

It is the shell structure caused by the effective mean field 
which produces the maxima that are observed experimentally in the addition 
energy $\mu (N+1)- \mu (N) = E_{N+1}-E_{N}+ e^2/C$ \cite{e2}.
Here $E_N$ is the single-particle energy of the effective
mean field in quantum dots, $e^2/C$  is the electrostatic energy and 
$\mu (N)$ is the 
chemical potential. The electrostatic energy is much larger
than the difference $E_{N+1}-E_{N}$, however, it is the fluctuations (shell
effects) of the difference that matters, at least for small 
quantum dots \cite{Leo}. Quantized electronic energies
have been observed even for nanometer--scale metal particles \cite{Tin}.
A similar effect is known in nuclear physics and for metallic clusters. 
There, shell effects due to single-particle motion create 
minima in the total potential energy surface which is dominated by the bulk 
energy, which is the classical liquid drop energy \cite{BM,NR,HB}.
The analogy goes further in that, in an isolated small 
quantum dot, the external magnetic 
field acts like the rotation on a nucleus thus creating new shell structure;
in this way superdeformation (axis ratio 2:1) has been established
for rotating nuclei owing to the shell gaps in the single-particle 
spectrum \cite{NR}.

In a previous study \cite{HeNa97} we have obtained various shapes of the 
quantum dot by energy minimization. In this context it is worth noting that 
at the particular values of the magnetic field, where a pronounced shell 
structure occurs, the energy minimum would be obtained for circular dots, 
if the particle number is chosen to be equal to the magic numbers. Deviations 
from those magic numbers usually give rise to deformed shapes at the energy 
minimum. To what extent these 'spontaneous' deformations actually occur 
(which is the Jahn--Teller effect \cite{JT}), is subject to 
more detailed experimental information. 
The far-infared spectroscopy in a small isolated quantum dot
could be a useful tool to provide pertinent data \cite{HeNa97}.

\vspace*{-0.1cm}
\begin{figure}
\epsfxsize=1.8in
\centerline{\epsffile{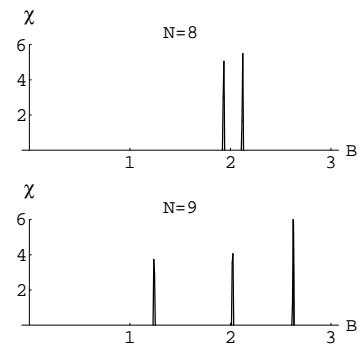}   }
\vglue 0.2cm
\caption{Magnetic susceptibility 
$\chi = -\partial ^2E_{{\rm tot}}/\partial B^2$ in arbitrary units
as a function of the magnetic field strength for the isotropic oscillator 
without $L^2$-term.}
\label{fig2}
\end{figure}

The question arises as to what extent our findings depend upon the particular
choice of the mean field. Here we confirm the qualitative argument
presented above that for sufficiently low electron numbers virtually any 
binding potential will produce the patterns found for the harmonic 
oscillator. We add to the Hamiltonian (\ref{ham}) the term 
$-\lambda \hbar \omega L^2$ where $L$ is the
dimensionless $z$-component of the angular momentum operator. In nuclear 
physics, the three dimensional analogue of the combined problem is known 
as the Nilsson model and has been quite successful in explaining the 
spectra of deformed nuclei \cite{BM}. For $\lambda >0$ the additional term 
lowers the energy levels of higher angular momenta, in other words, its 
effect mimics a bulging out of the lower part of the harmonic oscillator 
potential. As a consequence, it has the effect of interpolating between the
 oscillator and the square well single-particle spectrum \cite{NR}. 
 For $\omega _x\ne \omega _y$ and $\lambda \ne 0$ the combined 
Hamiltonian $H'=H - \lambda \hbar \omega L^2$
is non-integrable \cite{HeNa94} and the
level crossings encountered in Figs.1 become avoided level crossings. The
essential effect upon the lower end of the spectrum can be seen in the 
isotropic oscillator where the magnetic quantum number $m$ is a good quantum 
number. In this case $H'=H^{{\rm isotr}}- \lambda \hbar \omega m^2$. 

In Fig.1c we display a spectrum of such $H'$.
The shell structure, which prevails for 
$\lambda =0$ throughout 
the spectrum at $B\approx 1.23$ or $B\approx 2.01$, is now disturbed to an
increasing extent with increasing shell number. However, for the parameters 
chosen the structure is still clearly discernible for about seven shells, 
that is for particle numbers up to about twenty five. The lifting of
the degeneracies at $B=0$ is also clearly seen where the levels are split
according to the absolute values of $|m|$; it is this splitting which gives us
guidance in choosing an appropriate value for $\lambda $: for $B=0$ we aim at
levels that lie between the corresponding degenerate levels pertaining to the 
harmonic oscillator and the two dimensional square well where the splitting of
these levels is very strong.

\vspace*{-0.1cm}
\begin{figure}
\epsfxsize=1.8in
\centerline{\epsffile{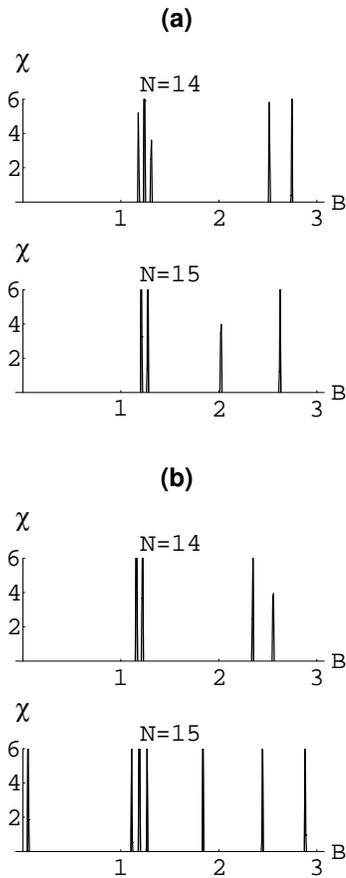}  }
\vglue 0.2cm
\caption{Similar to Fig.2 for different particle numbers and (a) for $\lambda
=0$ and (b) for $\lambda =0.01$.}
\label{fig3}
\end{figure}

When the magnetic field is changed continuously for a quantum dot of fixed
electron number, the ground state will undergo a rearrangement at the 
values of $B$, where level crossings occur. This should be observable in 
the magnetic susceptibility as it is proportional to the second derivative 
of the total energy with respect to the field strength. In Figs.2 and 3 
we have plotted 
$-\partial ^2E_{{\rm tot}}/\partial B^2$ where $E_{{\rm tot}}$ is the
sum of the single-particle energies filled from the bottom up to $N$ which 
is the electron number. We discern clearly distinct patterns depending on 
the electron number, in fact, the susceptibility appears to be a fingerprint 
pertaining to the electron number. Fig.2 and Fig.3a illustrate results for 
$\lambda =0$ and Fig.3b for $\lambda =0.01$. 
The deformed oscillator does not produce new features except for the 
fact that all lines in Figs.2 and 3 would be shifted towards the left in 
accordance with the discussion above.

All features of Figs.2 and 3 can be understood from the single-particle 
spectra displayed in Fig.1a and Fig.1c. If there is no level crossing, the 
second derivative of $E_{{\rm tot}}$ is a smooth function. The crossing of 
two occupied levels does not change the smoothness. In contrast, if the last 
occupied level takes
part in a level crossing in such a way that it is ascending before and
descending after the crossing, the second derivative of $E_{{\rm tot}}$ must
show a spike. In other words, spikes occur if the last occupied level is bent
concavely when viewed from the abscissa; a convex curvature has no effect. 
As a consequence, we understand the even-odd effect when comparing $N=8$ 
with $N=9$ in Fig.2 and $N=14$ with $N=15$ in Fig.3. The spin splitting 
caused by the magnetic field at $B\approx 2.01$ for $N=8$
is absent for $N=9$.

\vspace*{-0.1cm}
\begin{figure}
\epsfxsize=2.8in
\centerline{\epsffile{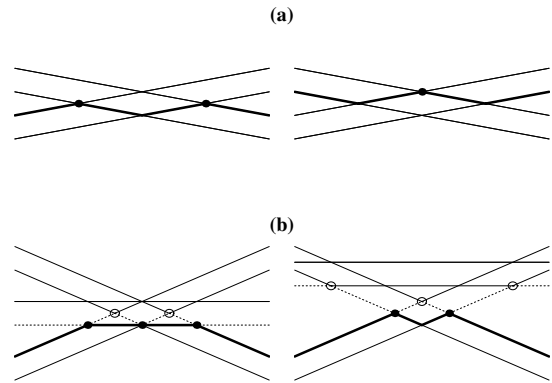}  }
\vglue 0.2cm
\caption{Blow ups of the relevant level crossings explaining (a) the features
in Fig.2 and (b) those of Fig.3. The left and right hand drawing of (a) 
refers to $N=8$ and $N=9$, respectively. The left hand and right hand drawing
of (b) refers to $\lambda =0$ and $\lambda =0.01$, respectively; here the 
thick and dotted lines refer to $N=14$ and $N=15$.} 
\label{fig4}
\end{figure}

 This becomes evident when looking at a blow up of
this particular level crossing which is illustrated in Fig.4a, where
the last occupied level is indicated as a thick line and the points where a 
spike occurs are indicated by a dot. In this way, we also understand that a 
splitting occurs for $N=6$ at $B\approx 1.23$ (not presented here) and no 
spike at all for $N=8$ at that same field strength. Note that 
the splitting is proportional to the effective spin magnetic moment $\mu ^*$.
Another illustration of
the even-odd effect, which persists through all numbers, is given in 
Fig.3. Now we obtain a triple splitting for $N=14$ at $B\approx
1.23$ which becomes a double splitting for $N=15$, while the double splitting
for $N=14$ at $B\approx 2.61$ becomes a single line for $N=15$.
The double splitting is associated with the crossing of only two double lines,
while the triple splitting originates from the crossing of three double lines.
The line distance at $B\approx 2.61$ is increased as it occurs at a larger 
field strength.

When we consider the crossing of only two double lines as is the case for 
$N=8$ and $N=9$, or $N=14$ and $N=15$ at $B\approx 2.61$, the effect of the 
$L^2$-term is unimportant. However, the $L^2$-term does become significant when 
we turn to level crossings of three or more double lines, in fact, the term 
removes the artificial pattern of level crossings of three and more lines.
We focus our attention at the multiple crossings around $B=1.2$ by comparing 
Fig.3a with Fig.3b. In Fig.4b we illustrate a blow up of the situation which 
produces the different results for $\lambda =0$ and $\lambda =0.01$. 
For $\lambda=0.01$ the levels with a non-zero slope are lowered as they refer 
to $m\ne 0$ while the horizontal levels remain unaffected,
hence the crossings change as seen by comparing the left and right hand drawing 
of Fig.4b. The last occupied level is indicated by a thick line and a dotted 
line for the $N=14$ and $N=15$ system, respectively; the corresponding 
positions, where spikes occur, are marked by solid dots and open circles. 
The even-odd effect is encountered again but appears converted for 
$\lambda =0.01$. Note, however, that the triple line for $N=15$ at $B=1.2$ is 
not a spin splitting but an effect due to the $L^2$-term; in other words, for 
$\mu ^*\to 0$ the double line distance of $N=14$ in Fig.3b would vanish 
while the triple line would not as it depends on $\lambda $.
Spin flips can also be understood by the same token, here the general rule
applies: the crossing of the top (bottom) with the bottom (top) line of a
double line causes the spin to flip. Hence, both lines of the double splitting 
in Fig.2 are associated with a spin flip ($N=8$), but neither of the single 
lines ($N=9$). Also, in Fig.3b the triplet ($N=15$) has no spin flip but if
$\lambda $ would be chosen sufficiently small, the central line of the triplet
would be split further and a spin flip would be associated with the doublet
(which is then part of a quartet).
Strictly speaking, the spikes are $\delta$-functions with a factor which is
determined by the angle at which the two relevant lines cross. Our figures are
numerical results which do not exactly reflect this feature. If the level
crossings are replaced by avoided crossings (Landau-Zener crossings), the 
lines would be broadened. This would be the case in the present model 
for $\lambda >0$ {\em and} $\beta >0$. Finite temperature will 
also result in line broadening. 

To summarize:
the consequences of shell structure effects for the addition energy of a small 
isolated quantum dot have been analyzed. At certain values of the magnetic
field strength shell structures appear in a quantum dot, also in cases where 
deformation does not give rise to magic numbers at zero field strength.
Measurements of the magnetic susceptibility are expected to reflect the 
properties of the single-particle spectrum and should display characteristic 
patterns depending on the particle number.
The latter property could be of interest in applications 
because it enables control of the electron number in small isolated
quantum dots and the magnetic field strength. 
The splittings due to the spin magnetic moment provide for a 
quantitative assessment of the effective spin magnetic moment $\mu^*$.

\vskip 0.5cm

The authors gratefully acknowledge illuminating discussions with
Anupam Garg; fruitful conversations with Yoram Alhassid, Dmitri Averin,
Aurel Bulgac and Gregor Hackenbroich are much appreciated.
We thank the Department of Energy  for
its support and the Institute for Nuclear Theory at the University of Washington 
for its congenial atmosphere.

\medskip
\noindent
heiss@physnet.phys.wits.ac.za, rashid@thsun1.jinr.dubna.su


\begin{references}
\bibitem{Kas} M.A. Kastner, Phys.Today {\bf 46}, 24 (1993);
R.C. Ashoori, Nature (London) {\bf 379}, 413 (1996) 
\bibitem{BM} A. Bohr and B.R. Mottelson, {\it Nuclear Structure}
(Benjamin, New York, 1975) Vol.2
\bibitem{NR} S.G. Nilsson and I. Ragnarsson, {\it Shapes and Shells in
Nuclear Structure} (Cambridge University Press, 1995)
\bibitem{HB} W.A. de Heer, Rev.Mod.Phys. {\bf 65}, 611 (1993);
M. Brack, {\it ibid} {\bf 65}, 677 (1993)
\bibitem{e1} D.J. Lockwood {\it et al}, Phys.Rev.Lett. {\bf 77}, 354 (1996)
\bibitem{e2} S.Tarucha {\it et al}, Phys.Rev.Lett. {\bf 77}, 3613 (1996)
\bibitem{e3} M. Fricke {\it et al}, Europ.Lett. {\bf 36}, 197 (1996)
\bibitem{DE} M. Dineykhan and G.V.Efimov, Reports Math.Phys. {\bf 36}, 287
(1995)
\bibitem{HK} D. Heitmann and J. Kotthaus, Phys.Today {\bf 46}, 56 (1993)
\bibitem{DN} M. Dineykhan and R.G. Nazmitdinov, Phys.Rev.{\bf B55}, 
13707 (1997); cond-mat$/9704202$
\bibitem{HeNa97} W.D. Heiss and R.G. Nazmitdinov, 
Phys.Lett. {\bf A222}, 309 (1996); Phys. Rev. {\bf B55}, N 20 (1997);
cond-mat$/9704216$
\bibitem{HHW}  G.Hackenbroich, W.D. Heiss and H.A. Weidenm\"uller,
Phys.Rev.Lett., to appear; cond-mat$/9702184$
\bibitem{HeNa94} W.D. Heiss and R.G. Nazmitdinov, Phys.Rev.Lett. {\bf 73}, 
1235 (1994)
\bibitem{Leo} L.P. Kouwenhoven {\it et al}, Z. Phys. {\bf B85}, 367 (1991)
\bibitem{Tin} C.T. Black, D.C. Ralph and M. Tinkham, Phys.Rev.Lett. {\bf 74},
3241 (1995); {\it ibid} {\bf 76}, 688 (1996); cond-mat/9701081 
\bibitem{JT} H.A. Jahn and E. Teller, Proc.R.Soc.London, Sec.A {\bf 161}, 220
(1937)
\end{references}
\end{document}